\documentclass[5p]{elsarticle}

\usepackage[utf8]{inputenc}
\usepackage[T1]{fontenc}

\usepackage{dblfloatfix}
\usepackage{graphicx}
\usepackage{mhchem}
\usepackage[font=footnotesize,labelfont=bf]{caption}
\usepackage[font=footnotesize,labelfont=bf]{subcaption}
\usepackage{booktabs}
\usepackage{amssymb}
\usepackage{multirow}
\usepackage{color}
\usepackage[normalem]{ulem}

\newcommand{\lambdabar}{{\mkern0.75mu\mathchar '26\mkern -9.75mu\lambda}}

\usepackage{dcolumn,booktabs}
\newcolumntype{d}[1]{D{.}{.}{#1}}
\newcommand\mc[1]{\multicolumn{1}{c}{#1}} % handy shortcut macro

\newcommand{\level}[3]{\ensuremath{#1,\,#2^{#3}}}
\newcommand{\levelnot}{\level{E_x}{J}{\pi}}

\journal{Physics Letters B}

\begin{document}

\begin{frontmatter}

\title{Experimental study of the $^{11}\text{B}(p,3\alpha)\gamma$ reaction at $E_p=0.5$--2.7 MeV}

\author[label2]{O.~S.~Kirsebom}

\author[label1]{A.~M.~Howard}

\author[label1]{M.~Munch}

\author[label1]{S.~Sablok}

\author[label1]{J.~A.~Swartz}

\author[label1]{H.~O.~U.~Fynbo\corref{cor1}}

\cortext[cor1]{Corresponding author}
\ead{fynbo@phys.au.dk}

\address[label2]{Institute for Big Data Analytics, Dalhousie University, Halifax, Nova Scotia B3H 4R2, Canada}

\address[label1]{Department of Physics and Astronomy, Aarhus University, DK-8000 Aarhus C, Denmark}

\begin{abstract}
Our understanding of the low-lying resonance structure in $^{12}$C remains incomplete. 
We have used the $^{11}\text{B}(p,3\alpha)\gamma$ reaction at proton energies of 
$E_p=0.5$--2.7 MeV as a selective probe of the excitation region above the $3\alpha$ 
threshold in $^{12}$C. Transitions to individual levels in $^{12}$C were identified 
by measuring the 3$\alpha$ final state with a compact array of charged-particle 
detectors. Previously identified transitions to narrow levels were confirmed and new 
transitions to broader levels were observed for the first time. Here, we report cross 
sections, deduce partial $\gamma$-decay widths and discuss the relative importance of 
direct and resonant capture mechanisms. 
\end{abstract}


\end{frontmatter}

\section{Introduction}\label{sec:intro}

The $p{}+^{11}\text{B}$ reaction has been extensively used to study the excitation structure of the $^{12}$C nucleus. This includes measurement of proton widths $\Gamma_p$, the partial $\gamma$ widths $\Gamma_{\gamma_0}$ and $\Gamma_{\gamma_1}$ to the two lowest levels in $^{12}$C, and the partial $\alpha$ widths $\Gamma_{\alpha_0}$ and $\Gamma_{\alpha_1}$ to the two lowest levels in $^{8}$Be~\cite{symons1963,segel1965,Becker:1987fk}. %
The focus of the present work are the two isospin $T=1$ resonances occurring at proton energies of $E_p=2.00$~MeV and $2.64$~MeV which correspond to the levels \level{17.76}{0}{+} and \level{18.85}{3}{-}.\footnote{Throughout this paper the notation {\levelnot} is used to denote excited nuclear levels, $E_x$ being the excitation energy in MeV and $J^{\pi}$ the spin and parity.} %
The $\gamma$ decay of these levels to lower-lying, unbound levels in $^{12}$C was studied by Hanna {\it et al.}~\cite{hanna1982} who identified two rather strong transitions feeding two narrow levels above the $3\alpha$ threshold: \level{17.76}{0}{+}$\rightarrow\;$\level{12.71}{1}{+} and \level{18.35}{3}{-}$\rightarrow\;$\level{9.64}{3}{-}. 

Using the conventional approach of detecting the $\gamma$ transitions with a large scintillator, Hanna {\it et al.}\ could not have identified weak transitions or transitions to broad levels. %
Recently, such transitions have been studied using a technique where the final level is identified by measuring the momenta of the three $\alpha$ particles resulting from its breakup~\cite{NIM_alcorta,kirsebom09_plb,laursen2016_2}. %
Here we wish to explore, first, if $\gamma$ transitions from the levels \level{17.76}{0}{+} and \level{18.85}{3}{-} to broad, lower-lying levels similar to those observed in Ref.~\cite{laursen2016_2} can be identified, and second, if the strength of the transitions already observed by Hanna {\it et al.}\ can be confirmed with this indirect detection method.

\section{Experiment}\label{sec:exp}

The experiment was performed at the 5~MV Van der Graaf accelerator at the Department of Physics and Astronomy at Aarhus University. The proton beam was directed on the target using electrostatic deflection plates and a magnetic bending stage. The beam size was defined by two variable apertures placed after the magnet both set at a separation of 2\,mm and placed 0.5\,m apart. 

The ion energy was adjusted by means of a generating voltmeter, which was calibrated on an absolute scale using the $^{27}\text{Al}(p,\alpha)^{24}\text{Mg}$ and $^{27}\text{Al}(p,\gamma)^{28}\text{Si}$ reactions. The energy spread of the beam was less than 1\,keV. Beam intensities of several 100\,nA can be delivered by the accelerator, but only beams of less than 1~nA were used for the experiment discussed here. The beam current was measured by a Faraday cup placed in a 1\,m long beam pipe downstream of the target chamber, specially designed to reduce the amount of beam back-scattered from the Faraday cup to the detector setup.     

Long measurements were performed at proton energies of $E_p=2.00$~MeV and $2.64$~MeV. At the lower energy, a total of 295~$\mu$C was directed on the target over a period of 211~hours, which corresponds to an average current of 0.39~nA. For the higher energy setting, the corresponding numbers are 124~$\mu$C, 77~hours, and 0.45~nA. Additionally, multiple, short measurements were performed across the energy range 0.5--3.5~MeV as reported in Ref.~\cite{munch2020}.

%\noindent $E_p = 2.00$~MeV, $IC = 295\, \mu$C, 211 hours, $DT=8.8\%$, average current 0.39nA \\
%$E_p = 2.64$~MeV, $IC = 124\, \mu$C, 77 hours, $DT=11.7\%$, average current 0.45nA\\
%%$E_p = 2.37$~MeV, $IC = 14.5\, \mu$C, 6.0 hours\\

The target consisted of a layer of 12.6(1.2)~$\mu$g/cm$^2$ isotope-enriched $^{11}$B deposited on a 4~$\mu$g/cm$^2$ carbon backing~\cite{munch2020}. %
The target was manufactured from 99\% enriched $^{11}\text{B}$ by slow evaporation in a Cu crucible. In addition to B, C, and Cu, the target was found to contain H and O impurities, likely due to condensation of water vapor. The presence of these impurities was inferred from the corresponding Rutherford scattering peaks in the singles spectra. Considering the known target constituents, the only open three-body channels at the beam energies used in this study are $p+{}^{10}\text{B} \rightarrow 2\alpha + {}^3\text{He}$ and $p+{}^{11}\text{B} \rightarrow 3\alpha$.%

\begin{figure}[t!]
    \includegraphics[width=\columnwidth]{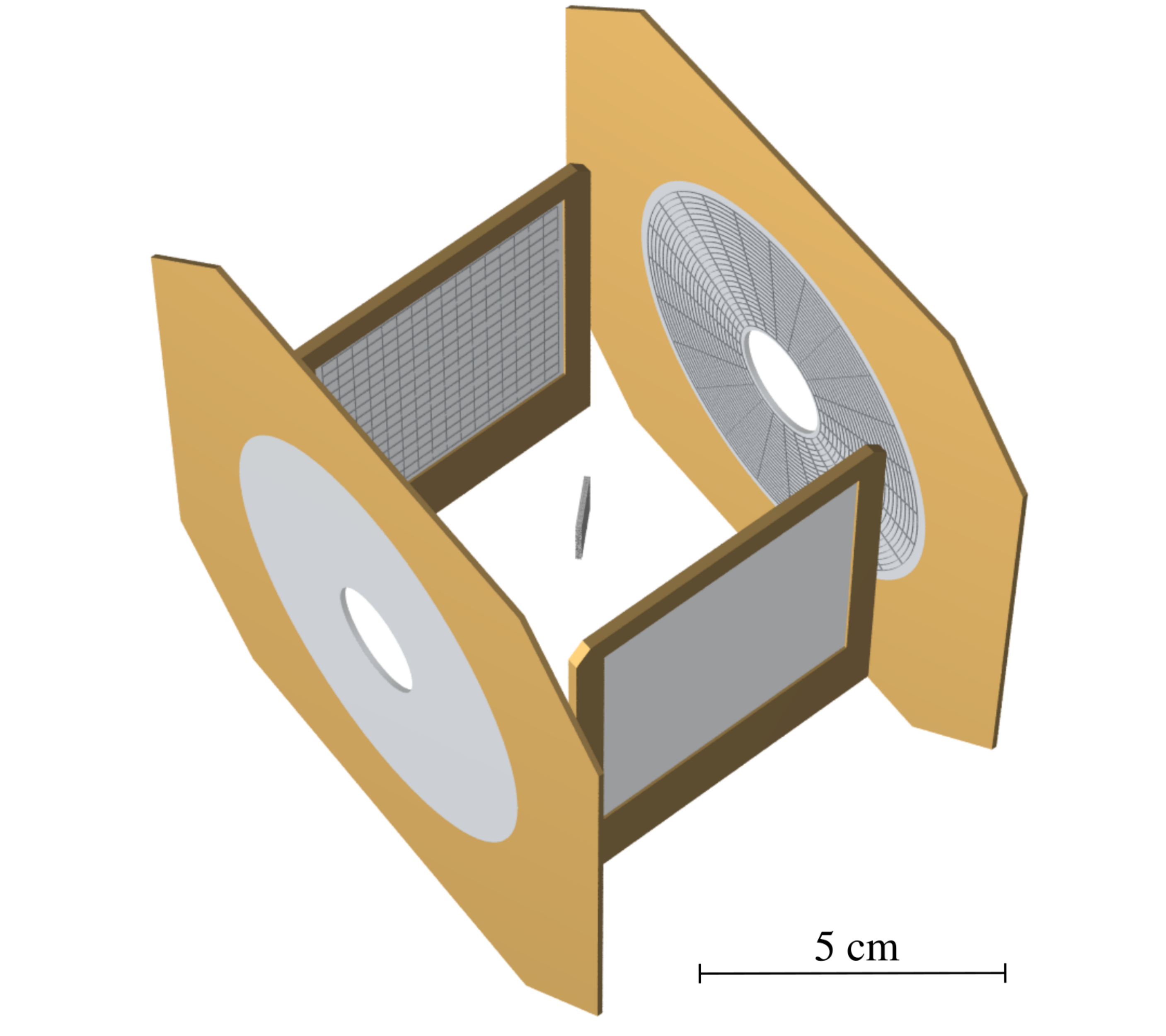}
    \caption{\label{fig:setup} Schematic illustration of the detector setup which consisted of two annular and two square double-sided silicon strip detectors. The proton beam enters the setup through one annular detector and exits through the other. The $^{11}\text{B}$ target is placed in the middle of the setup at a 45$^{\circ}$
    angle with respect to the beam axis.}
\end{figure}

The target was placed in the middle of a compact array of double sided Si strip detectors (DSSDs) at an angle of 45$^{\circ}$ with respect to the axis defined by the beam, as shown in Fig.~\ref{fig:setup}. %
Annular DSSDs with 24 ring strips and 32 annular strips were placed upstream and downstream of the target, and two square DSSDs with 16 horizontal strips and 16 vertical strips were placed on either side of the target orthogonal to the beam axis. 

The electronics and data acquisition consisted of a VME based system with ADCs and TDCs fed by signals from a chain of preamplifiers and amplifiers. 
The dead time was around 10\% with trigger rates of several kHz.

\section{Event selection}

The data is analyzed following an approach similar to that of Laursen {\it et al.}~\cite{laursen2016_2}. Particle energies and hit positions on the DSSDs are determined by requiring an energy difference of at most $\pm 50$~keV in front and back strips. Energy conservation cannot be used as a condition to reduce unwanted background because we are searching for events where some of the energy is carried away by a $\gamma$ ray. However, the momentum carried away by the $\gamma$ ray is sufficiently small that we can require momentum conservation of the three $\alpha$ particles. Hence, $3\alpha$ events are identified as triple coincidence events fulfilling both a TDC cut of $\pm 15$~ns and momentum conservation, but not necessarily energy conservation. 

\begin{figure}[h!]%
    \includegraphics[width=0.99\columnwidth,clip=true,trim=0 0 0 50]{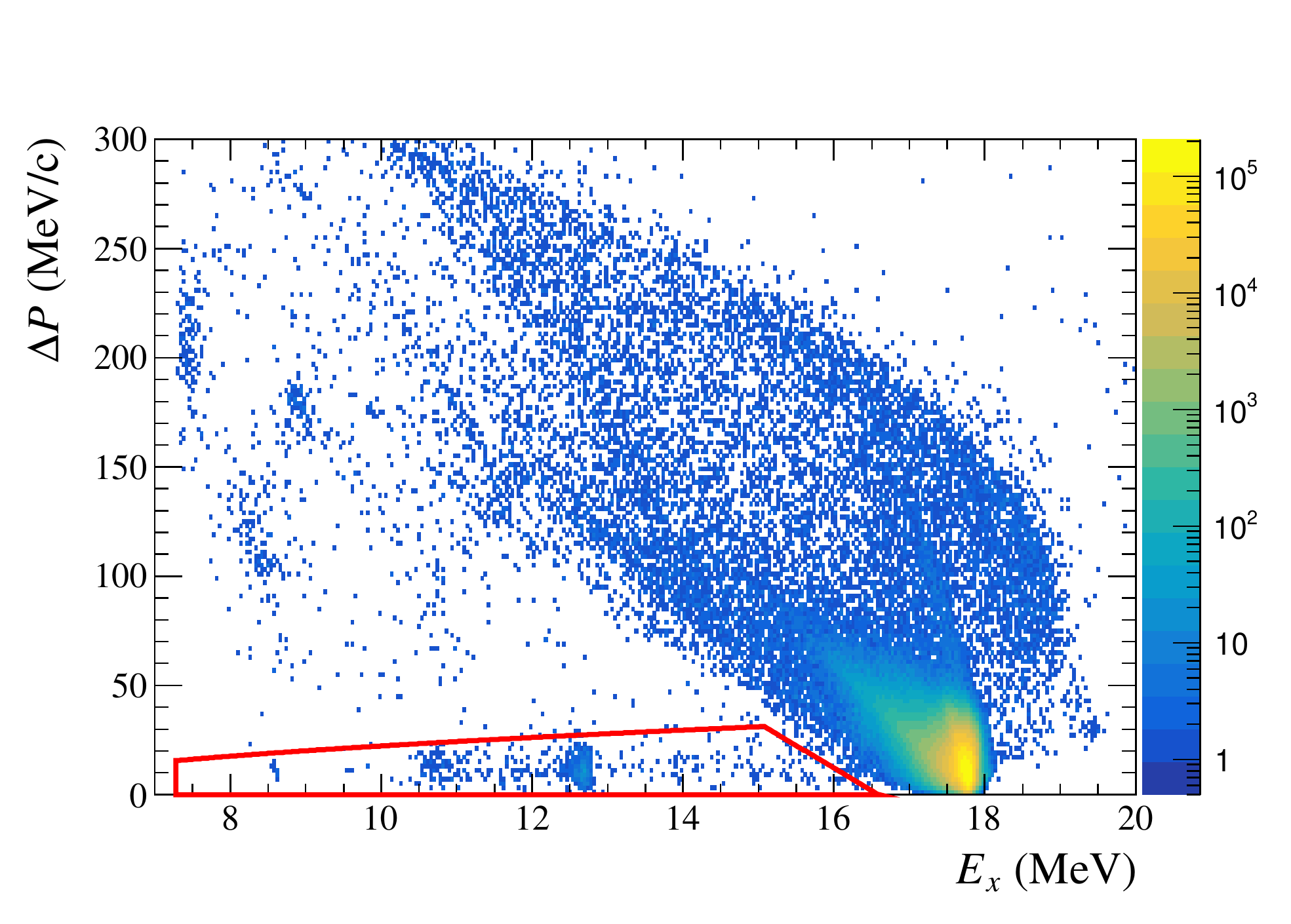}
    \caption{\label{fig:dPEx1} Triple-concidence data obtained at $E_p=2.00$~MeV. The $x$ axis is the excitation energy in $^{12}$C, and the $y$ axis is the total momentum in the centre of mass frame, both determined from the energies and positions of the three detected particles. The events enclosed by the red contour fulfill momentum conservation, but not energy conservation, and are therefore interpreted as $\gamma$-delayed 3$\alpha$ emissions from $^{12}$C.}
\end{figure}

\begin{figure}[h!]%
    \includegraphics[width=0.99\columnwidth,clip=true,trim=0 0 0 50]{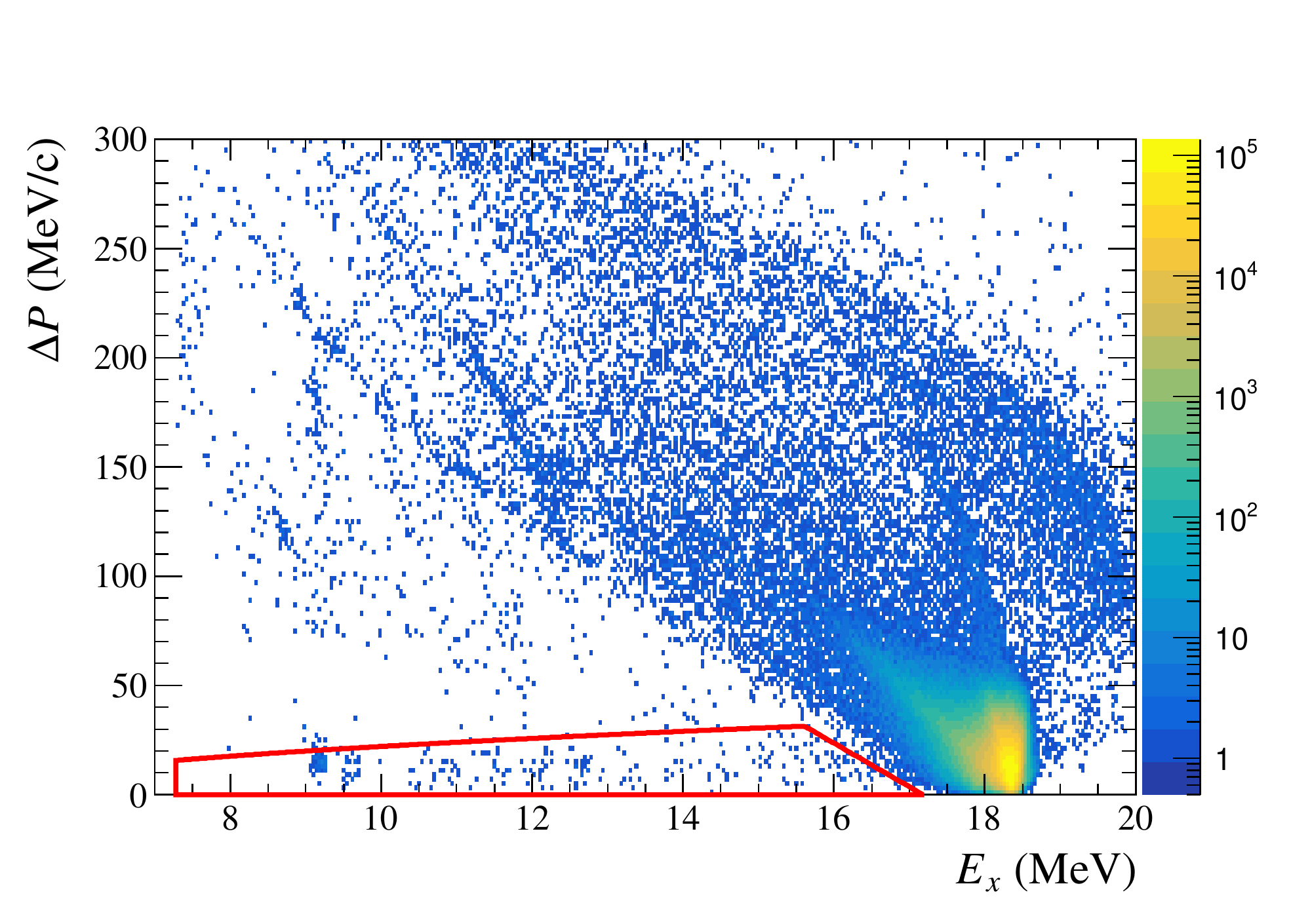}
    \caption{\label{fig:dPEx2} Triple-concidence data obtained at $E_p=2.64$~MeV. The axes are the same as in Figure \ref{fig:dPEx1}.}
\end{figure}

Figures \ref{fig:dPEx1} and \ref{fig:dPEx2} show scatter plots of the total momentum in the centre of mass (c.m.) frame versus the $^{12}$C excitation energy calculated from the triple-coincidence events. 
The intense groups of events just below and just above $E_x = 18$~MeV in the two figures correspond to 3$\alpha$ decays directly from the levels \level{17.76}{0}{+} and \level{18.85}{3}{-}, respectively. These events fulfill both energy and momentum conservation. The events further to the left from these intense regions, enclosed by the red contours, are interpreted as events where some of the energy is carried away by a $\gamma$ ray, and they are therefore the events of interest. 

To assert that these events are in fact genuine triple-$\alpha$ coincidences as opposed to, say, two $\alpha$ particles in coincidence with a noise signal, the following checks were made: First, the energy distribution of the individual detections was inspected to ensure that the energies were comfortably above the ADC thresholds. Second, the spatial distribution of the detections across the surface of the DSSDs was inspected to verify that the events were not caused by a single or a few noisy strips. Third, the effect on the event rate of widening the TDC cut was studied. For genuine coincidences one expects the event rate to plateau once the width the TDC cut exceeds the experimental resolution, whereas for random coincidences one expects the event rate to continue increasing.

Looking at Figures \ref{fig:dPEx1} and \ref{fig:dPEx2} one notes the occurrence of a number of clusters of events at low excitation energy ($E_x \sim 7$--9~MeV) which exhibit a considerable momentum mismatch ($\Delta P \sim 50$--250~MeV/c). Each of these clusters was subject to a careful analysis, which revealed all but one of the clusters to be comprised of random coincidences, in most cases involving $p+{}^{11}$B coincidences or $p+p$ coincidences due to elastic scattering on H impurities in the target. 
A dedicated analysis followed to clarify the origin of the single cluster that could not be attributed to random coincidences. This cluster was found to be comprised of $p+p$ coicidences in which one of the protons, having penetrated into the active volume of one of the DSSDs, is backscattered into a second DSSD, thus producing two separate detections with a combined energy close to the original proton energy. Having identified $p+p$ and $p+{}^{11}$B coincidences as significant sources of background, dedicated kinematic cuts were implemented to selectively remove such events. As a result, the event density within the clusters was substantially reduced and some clusters were fully removed, leaving only the few clusters visible in Figures \ref{fig:dPEx1} and \ref{fig:dPEx2}.%

\begin{figure}[t!]
    \includegraphics[width=0.99\columnwidth,clip=true,trim=0 50 0 50]{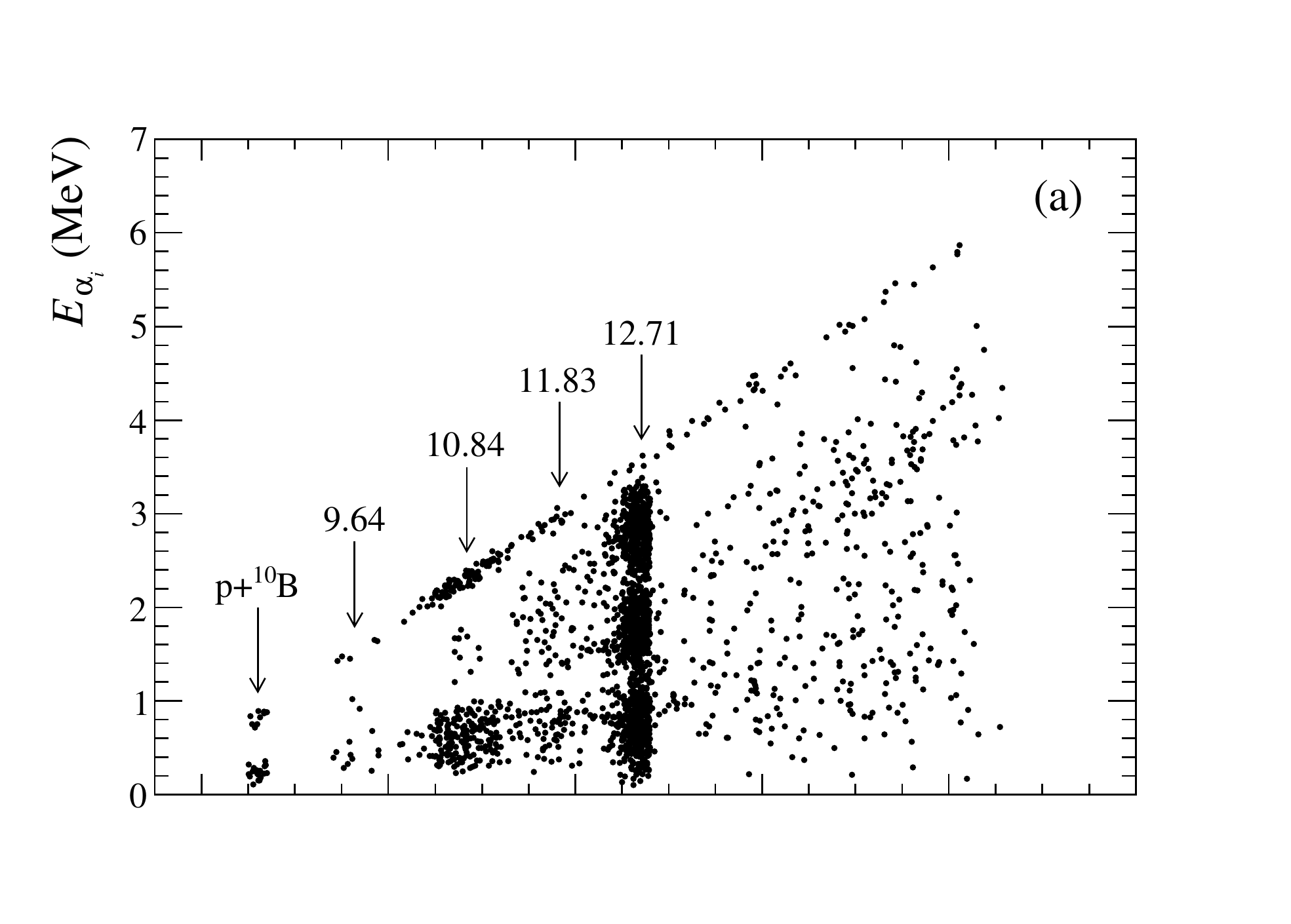}
    \includegraphics[width=0.99\columnwidth,clip=true,trim=0 0 0 40]{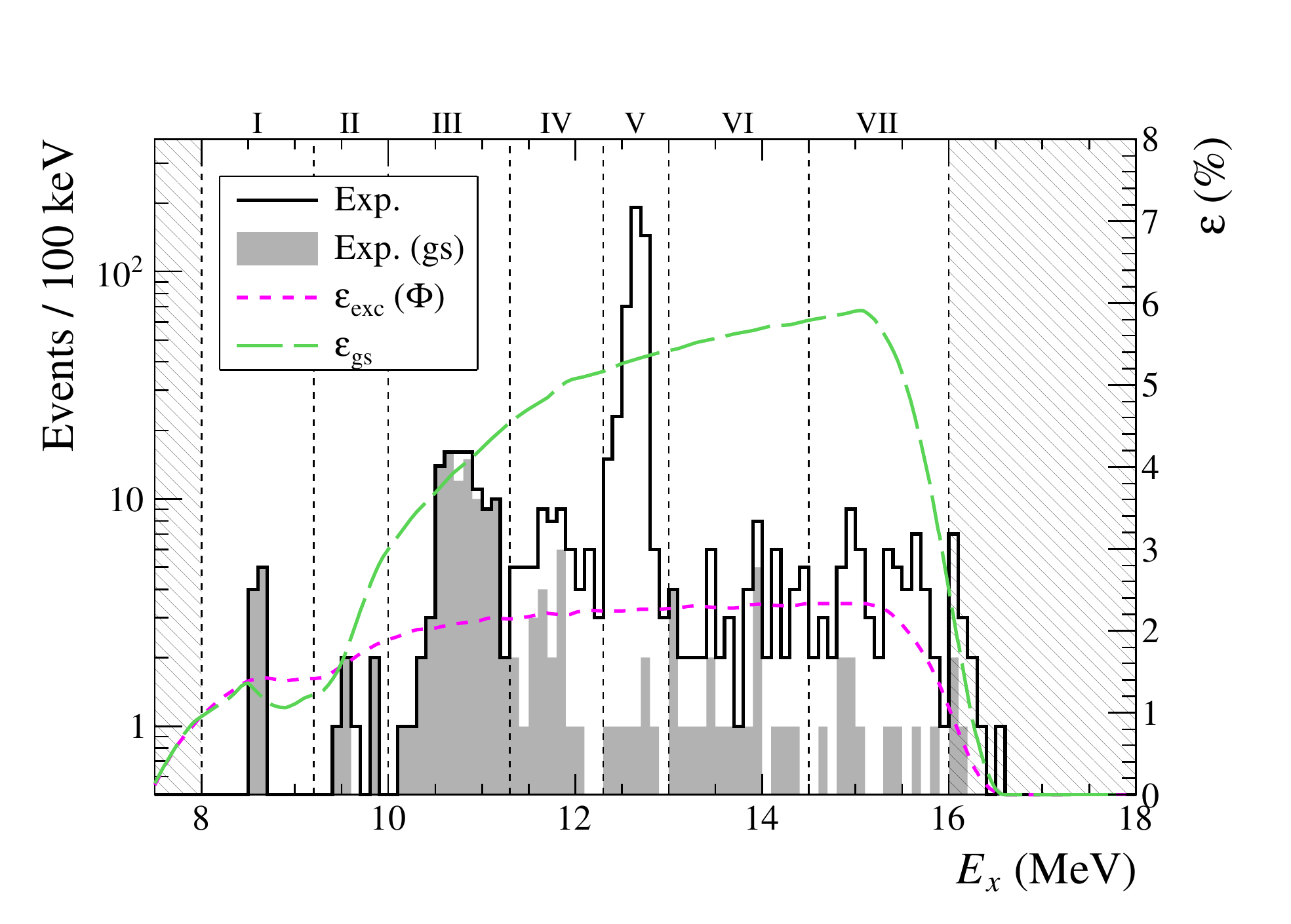}
    \caption{\label{fig:2MeV} $\gamma$-delayed $3\alpha$ spectra obtained at $E_p=2.00$~MeV. Panel (a) is a scatter plot of the excitation energy in $^{12}$C versus the individual energies of the three detected $\alpha$ particles in the $^{12}$C rest frame. The positions of selected known levels in $^{12}$C are indicated. Panel (b) is the projection of the scatter plot on the excitation energy axis. The shaded histogram is obtained by selectively projecting events in which the $3\alpha$ breakup proceeds via the ground state of $^8$Be. The green and magenta (short- and long-dashed) curves show the detection efficiencies determined from Monte-Carlo simulations.}
\end{figure}

\begin{figure}[t!]
    \includegraphics[width=0.99\columnwidth,clip=true,trim=0 50 0 50]{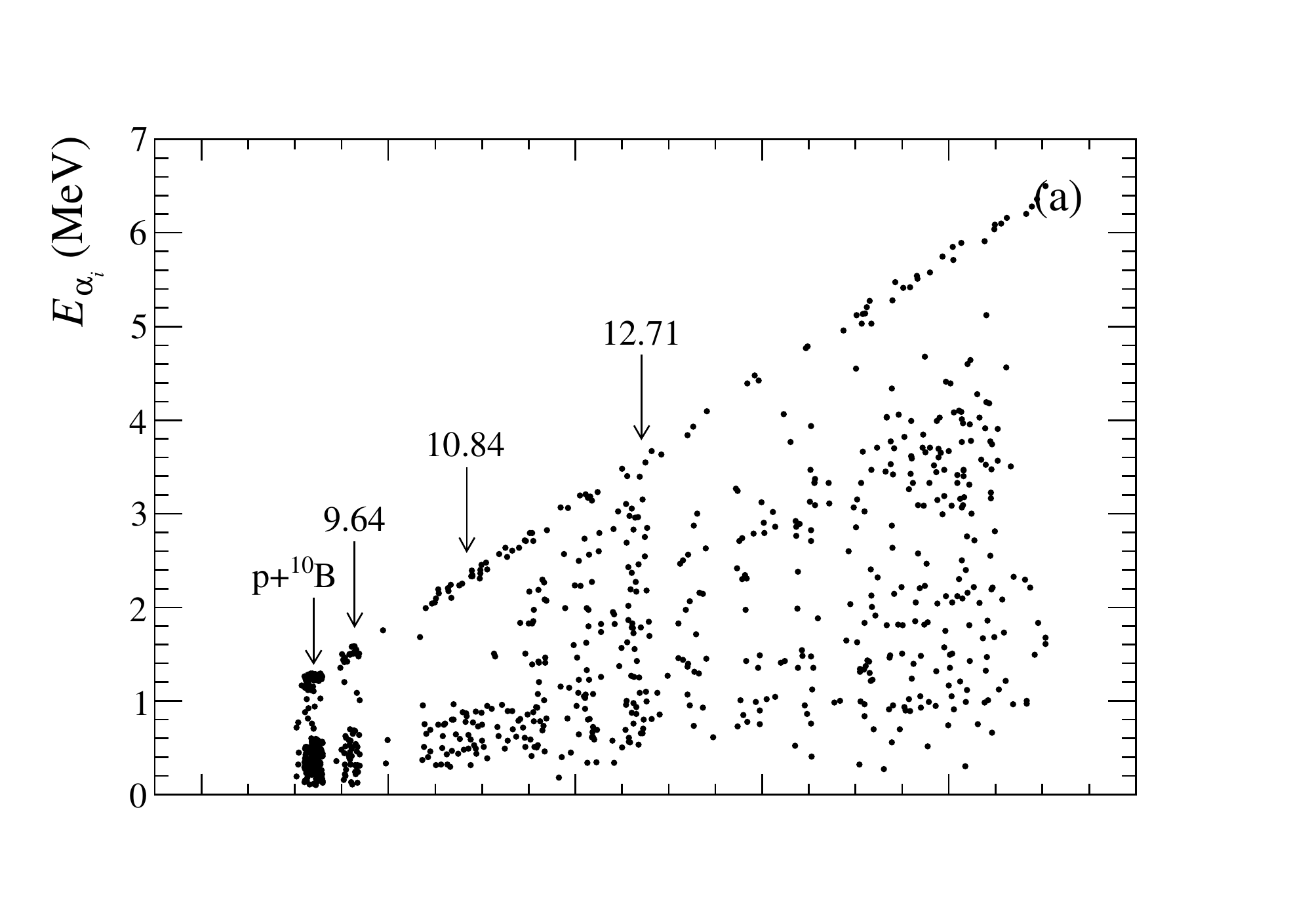}
    \includegraphics[width=0.99\columnwidth,clip=true,trim=0 0 0 40]{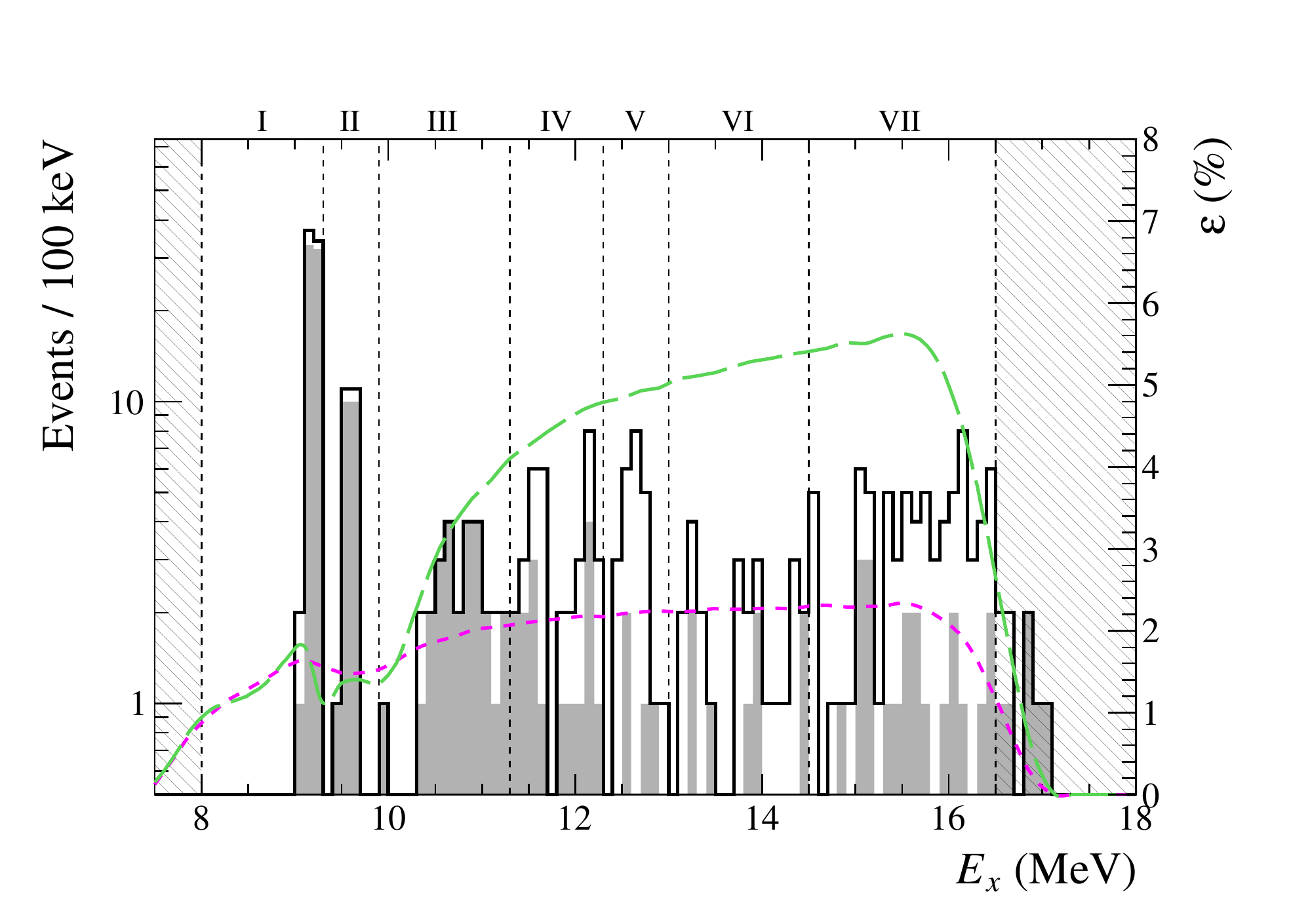}
    \caption{\label{fig:2.64MeV} Similar to Figure \ref{fig:2MeV} but for $E_p=2.64$~MeV.}
\end{figure}

Figures \ref{fig:2MeV} and \ref{fig:2.64MeV} focus specifically on those events fulfilling momentum conservation, but not energy conservation. The upper panels show scatter plots of the excitation energy in $^{12}$C versus the individual energies of the three $\alpha$-particles in the $^{12}$C rest frame. These scatter plots show the different $3\alpha$ breakup mechanisms of the levels in $^{12}$C populated in the $\gamma$ decays. The diagonal lines from the lower left to the upper right represent breakups that proceed by $\alpha$ decay to the ground state of $^8$Be. Owing to parity and angular momentum conservation this decay mechanism is only allowed for natural-parity levels in $^{12}$C. The two $\alpha$ particles from the subsequent breakup of $^8$Be, detected in coincidence with the primary $\alpha$ particle, form a broad band running from left to right with half the slope of the upper diagonal. The positions of known levels in $^{12}$C are indicated on the scatter plots.

The lower panels of Figures \ref{fig:2MeV} and \ref{fig:2.64MeV} show the projections of the scatter plots on the excitation energy axes with the shaded histograms providing the projection selectively for the events on the diagonals, which fulfill the condition $E_{2\alpha}<210$~keV for at least one pair of $\alpha$ particles, $E_{2\alpha}$ being the relative kinetic energy of the pair. The coloured curves on these plots will be discussed later. From the trigger rate and the width of the TDC gate we estimate the number of random coincidences to be 6 events in Figure \ref{fig:2MeV} and 9 events in Figure \ref{fig:2.64MeV}.

The transitions identified by Hanna {\it et al.} \cite{hanna82}, namely \level{17.76}{0}{+}$\rightarrow\;$\level{12.71}{1}{+} and \level{18.35}{3}{-}$\rightarrow\;$\level{9.64}{3}{-}, are clearly identifiable in Figure \ref{fig:2MeV} and \ref{fig:2.64MeV}, respectively. In addition to these, there is clear evidence for additional transitions. These will be discussed further later.

\section{Cross sections}

We determine the capture cross section, $\sigma_{\gamma}$, from the number of observed events in each excitation energy bin, taking into account the triple-$\alpha$ detection efficiency, the target thickness, the integrated charge on the target, and the dead-time of the data acquisition system. 
The cross sections thus obtained at $E_p=2.00$~MeV and $2.64$~MeV are summarized in Table \ref{tbl:tab3} and \ref{tbl:tab4}, respectively.

\begin{table}[b]
\setlength{\tabcolsep}{4pt}
{\footnotesize
\centering
\caption{$^{11}\text{B}(p,3\alpha)\gamma$ cross section at $E_p=2.00$~MeV. $E_x$ is the $^{12}$C excitation energy inferred from the momenta of the 
three $\alpha$ particles; $\sigma_{\gamma}$ is the cross section and is subject to an additional 10\% systematic uncertainty from the target thickness; 
the events are divided into two groups: those that correspond to breakups proceeding via the $^8$Be ground state (gs) and those that do not (exc). The first energy bin (I) is not included since all the events in this bin are attributed to $p+{}^{10}$B.}
\label{tbl:tab3}
\begin{tabular*}{\linewidth}{@{\extracolsep{\fill}}cc *{1}{d{1.6}} *{2}{d{2.6}} *{1}{d{1.6}}}
\toprule 
\multirow{2}{*}{Bin}  &  \multirow{2}{*}{$E_x$~(MeV)}  &  \multicolumn{4}{c}{$\sigma_{\gamma}$~($\mu$b)} \\ \cline{3-6}
 & & \mc{gs} & \mc{exc} & \mc{tot} & \mc{Ref.~\cite{hanna1982}}\\
    \midrule
II  &  9.2--10.0  &  0.18(9)   &  \mc{0.005--0.11}  &  0.24(10)  &   \\
III & 10.0--11.3  &  1.56(27)  &  0.21(11)          &  1.77(29)  &   \\ 
IV  & 11.3--12.3  &  0.27(8)   &  1.6(5)            &  1.8(5)    &   \\
V   & 12.3--13.0  &  0.09(4)   &  13.9(26)          &  14.0(26)  &  6.5(26) \\
VI  & 13.0--14.5  &  0.25(7)   &  0.9(4)            &  1.2(4)    &   \\
VII & 14.5--16.0  &  0.13(5)   &  1.6(5)            &  1.8(5)    &   \\
\bottomrule   
\end{tabular*}
}
\end{table}

\begin{table}[h]
\setlength{\tabcolsep}{5pt}
{\footnotesize
\caption{$^{11}\text{B}(p,3\alpha)\gamma$ cross section at $E_p=2.64$~MeV. See caption of Table~\ref{tbl:tab3} for further information.}
\label{tbl:tab4}
\centering
\begin{tabular*}{\linewidth}{@{\extracolsep{\fill}}cc *{1}{d{1.6}} *{3}{d{1.5}} }
\toprule 
\multirow{2}{*}{Bin}  &  \multirow{2}{*}{$E_x$~(MeV)}  &  \multicolumn{4}{c}{$\sigma_{\gamma}$~($\mu$b)} \\ \cline{3-6}
 & & \mc{gs} & \mc{exc} & \mc{tot} & \mc{Ref.~\cite{hanna1982}}\\
    \midrule
II  &  9.3--9.9   &  2.4(7)    &  \mc{0.08--0.5}  &  2.7(7)  &  4.2(17)      \\
III &  9.9--11.3  &  1.3(4)    &  \mc{0.06--0.4}  &  1.6(4)  &  \\ 
IV  & 11.3--12.3  &  0.59(18)  &  1.6(7)          &  2.1(7)  &  \\
V   & 12.3--13.0  &  0.15(7)   &  1.6(6)          &  1.7(6)  &  \\
VI  & 13.0--14.5  &  0.26(11)  &  1.3(5)          &  1.5(6)  &  \\
VII & 14.5--16.5  &  0.71(19)  &  4.5(15)         &  5.2(15) &  \\
\bottomrule   
\end{tabular*}
}
\end{table}

The detection efficiency depends on the 3$\alpha$ breakup mechanism and differs significantly between breakups that do and do not proceed via the ground state (g.s.) of $^8$Be. The green and magenta (short- and long-dashed) curves in the lower panels of Figures \ref{fig:2MeV} and \ref{fig:2.64MeV} show the detection efficiencies determined from Monte-Carlo simulations. %
For the excited channel, phase-space ($\Phi$) simulations were used to estimate the detection efficiency in all excitation energy bins, except the bins containing the \level{11.83}{2}{-} and \level{12.71}{1}{+} levels where more accurate models~\cite{fynbo03} were used. The error resulting from adopting the phase-space approximation is estimated to be at most $\sim 15$\%, which we include as an additional uncertainy on the detection efficiency for those energy bins where phase-space simulations were used. For the other bins, and for the g.s.\ channel where the angular distributions of Ref.~\cite{munch2020} were used, we adopt a 5\% model uncertainty.

We note that the ratio of triple-coincidence events to single events predicted by the simulation for the g.s.\ channel is 15\% below the experimental ratio. We ascribe this to inaccuracies in the representation of the beam-target-detector geometry in the simulation and account for it by including an additional 15\% uncertainty on our efficiency estimate.
We find the detection efficiency to be insensitive to uncertainties in the ADC thresholds, except for the lowest excitation energy bin ($E_x < 9.2$~MeV) where ADC thresholds contribute an estimated 8\% to the overall uncertainty. 
These uncertainty contributions are all added up in quadrature, and finally added linearly with the statistical counting uncertainty to obtain the overall uncertainty on the cross section in each excitation energy bin.

\section{Deduced $\gamma$-ray widths}

The excitation functions of the $\gamma$ rays to the \level{9.64}{3}{-} and \level{12.71}{1}{+} levels have been measured in considerable detail in the energy range $E_p = 1.8$--$3.0$~MeV by Hanna {\it et al.}~\cite{hanna1982} by means of conventional $\gamma$-ray spectroscopy. Both excitation functions were found to be resonant, allowing the authors to attribute the $\gamma$ rays to the transitions \level{17.76}{0}{+}$\rightarrow\;$\level{12.71}{1}{+} and \level{18.35}{3}{-}$\rightarrow\;$\level{9.64}{3}{-}, respectively. %
One drawback of the indirect experimental approach adopted in the present work, which involves detecting the three $\alpha$ particles rather than the $\gamma$ ray, is the reduced event rate compared to conventional $\gamma$-ray spectroscopy. Therefore, excitation functions could not be obtained in a reasonable amount of time and measurements were limited to a few selected beam energies. %
In the absence of excitation functions to support a resonant interpretation of the measured cross sections, we rely on the findings of Hanna {\it et al.}~\cite{hanna1982} concerning the resonant character of the $\gamma$ rays to the \level{9.64}{3}{-} and \level{12.71}{1}{+} levels, as well as theoretical estimates of the direct-capture cross section, to justify a resonant interpretation of the new $\gamma$ rays observed in this work. The theoretical estimates of direct-capture cross section will be discussed next. %

\subsection{Direct capture}

For the purpose of estimating the (E1) direct-capture capture cross section, we adopt the 
model of Rolfs~\cite{rolfs1973} which approximates the many-nucleon problem 
by a two-body problem in which the projectile and target are treated as inert cores 
and their interaction is described by a square-well potential with the depth adjusted 
to reproduce the binding energy of the final state. This simple model was found 
to yield accurate results for the capture reaction $^{16}\text{O}(p,\gamma)$ to 
the two bound levels in $^{17}$F, which both are well described by a simple, 
single-particle configurations involving only a single orbital~\cite{rolfs1973}. 

Here, we apply the model to capture transitions to levels in $^{12}$C which are 
not well described by single-particle configurations and also are unbound with 
respect to decay to the $3\alpha$ final state. Therefore, we do not expect 
the model to be very accurate and will use its predictions merely as 
order-of-magnitude estimates, accurate only within a factor of 2--3 or so.

\begin{figure}[h!]
    \includegraphics[width=0.99\columnwidth,clip=true,trim=0 0 0 0]{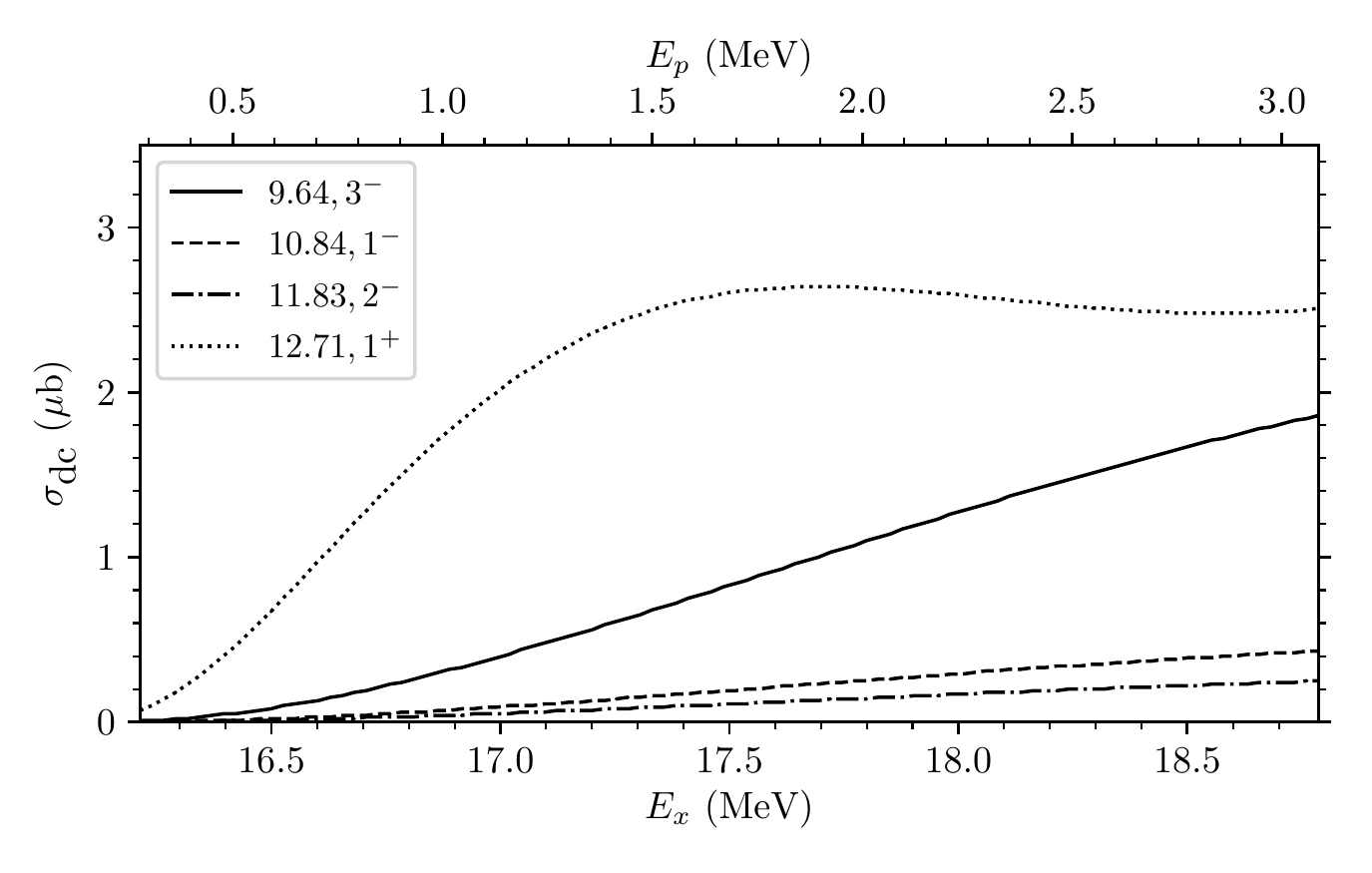}
    \caption{\label{fig:dc-cross-sec} Estimates of the cross section for $p+{}^{11}\text{B}$ direct capture to four selected levels in $^{12}$C based on the model of Ref.~\cite{rolfs1973}.}
\end{figure}

Estimates of the direct capture-cross section to four known levels 
in $^{12}$C computed with the model of Rolfs using the parameters 
listed in Table~\ref{tbl:rolfs}, are shown in Fig.~\ref{fig:dc-cross-sec}. %
The computed cross sections are proportional 
to the assumed spectroscopic factor, which is not predicted by the model 
itself. For the \level{12.71}{1}{+} level we take the spectroscopic factor 
from Ref.~\cite{adelberger77}. For the remaining levels we use 
the average values of the spectroscopic factors compiled in 
Ref.~\cite{tunl12}, noting that there is a substantial spread 
($\sim 50\%$) in the spectroscopic factors obtained by different 
authors. %
In all cases, we assume a single-orbital configuration, with 
$\ell_{\text{i}}=1$ for the \level{12.71}{1}{+} level and 
$\ell_{\text{i}}=2$ for the remaining levels. %
The channel radius was taken to be 4.38~fm. 
\begin{table}[b]
\setlength{\tabcolsep}{5pt}
\centering
{\footnotesize
\caption{Parameters used for estimating the cross section 
for direct capture to four levels in $^{12}$C based on 
the model of Ref.~\cite{rolfs1973}. $\ell_{\text{i}}$ are 
the orbital angular momenta in the entrance channel, 
$\ell_{\text{f}}$ is the orbital angular momentum 
assumed for the final state, and $S$ is the spectroscopic 
factor. The spectroscopic factor of the \level{12.71}{1}{+} 
level was taken from Ref.~\cite{adelberger77}; for the 
remaining levels we use the average values of the spectroscopic 
factors reported in Ref.~\cite{tunl12}. The channel radius was 
taken to be 4.38~fm.}
\label{tbl:rolfs}
\begin{tabular*}{0.55\linewidth}{@{\extracolsep{\fill}}ccccc}
\toprule 
\mc{$E_x$~(MeV)}  &  \mc{$J^{\pi}$}  & 
\mc{$\ell_{\text{i}}$}  &  \mc{$\ell_{\text{f}}$}  &
\mc{$S$}  \\ 
\midrule
 9.64  &  $3^-$  &  1,3  &  2  &  0.30    \\
10.84  &  $1^-$  &  1,3  &  2  &  0.23    \\ 
11.83  &  $2^-$  &  1,3  &  2  &  0.11    \\
12.71  &  $1^+$  &  0,2  &  1  &  0.86    \\
\bottomrule   
\end{tabular*}
}
\end{table}

%Reduction/increase in $r_0$ by 5\% leads to $\sim 25$\% decrease/increase in $\sigma_{\text{DC}}$ for lf=2 levels and 12\% decrease/increase for lf=1 level.

The excitation functions measured by Hanna {\it et al.}~\cite{hanna1982} 
(at 90$^{\circ}$) indicate that direct capture contributes at most $\sim 10\%$ 
to the total capture cross section to the \level{12.71}{1}{+} level at 
$E_p=2.00$~MeV, corresponding to 1.4~$\mu$b, which is within a factor of two of 
the cross section predicted by the model (2.6~$\mu$b). %
Similarly, the direct-capture contribution to the cross section to the \level{9.64}{3}{-} 
level can be estimated to be at most $\sim 15\%$ of the total capture cross section at 2.64 MeV, 
corresponding to 0.4~$\mu$b, a factor of four below the model prediction (1.6~$\mu$b). %
Thus, we conclude that our rather crude model provides reasonable estimates 
of the direct-capture cross section, with a tendency to overestimate the actual 
cross section by a factor of two to four. Comparing the predicted direct-capture 
cross sections (Fig.~\ref{fig:dc-cross-sec}) to the measured total capture cross 
sections (Tables~\ref{tbl:tab3} and \ref{tbl:tab4}), we conclude that resonant 
capture is likely to be the dominant mechanism in most energy bins, but with 
a substantial contribution from direct capture.

\subsection{Resonant capture}

The goal of the analysis is to calculate the partial $\gamma$ widths of the 
levels in $^{12}$C mediating the observed (resonant) capture transitions. For this we use 
the resonant cross section formula,
\begin{equation}\label{eq:resonant}
\sigma_{\gamma,\textrm{R}} = 4\pi \lambdabar^2 \omega \Gamma_p \Gamma_{\gamma} / \Gamma^2
\end{equation}
where $\omega = \tfrac{1}{8}(2J+1)$ is the spin statistical factor appropriate for $p+^{11}$B. 
Using this equation the partial $\gamma$ decay widths can be determined from the measured 
cross sections, provided the partial proton decay widths ($\Gamma_p$) and the total widths 
($\Gamma$) are known.  

\begin{figure}[t]
    \includegraphics[width=0.99\columnwidth]{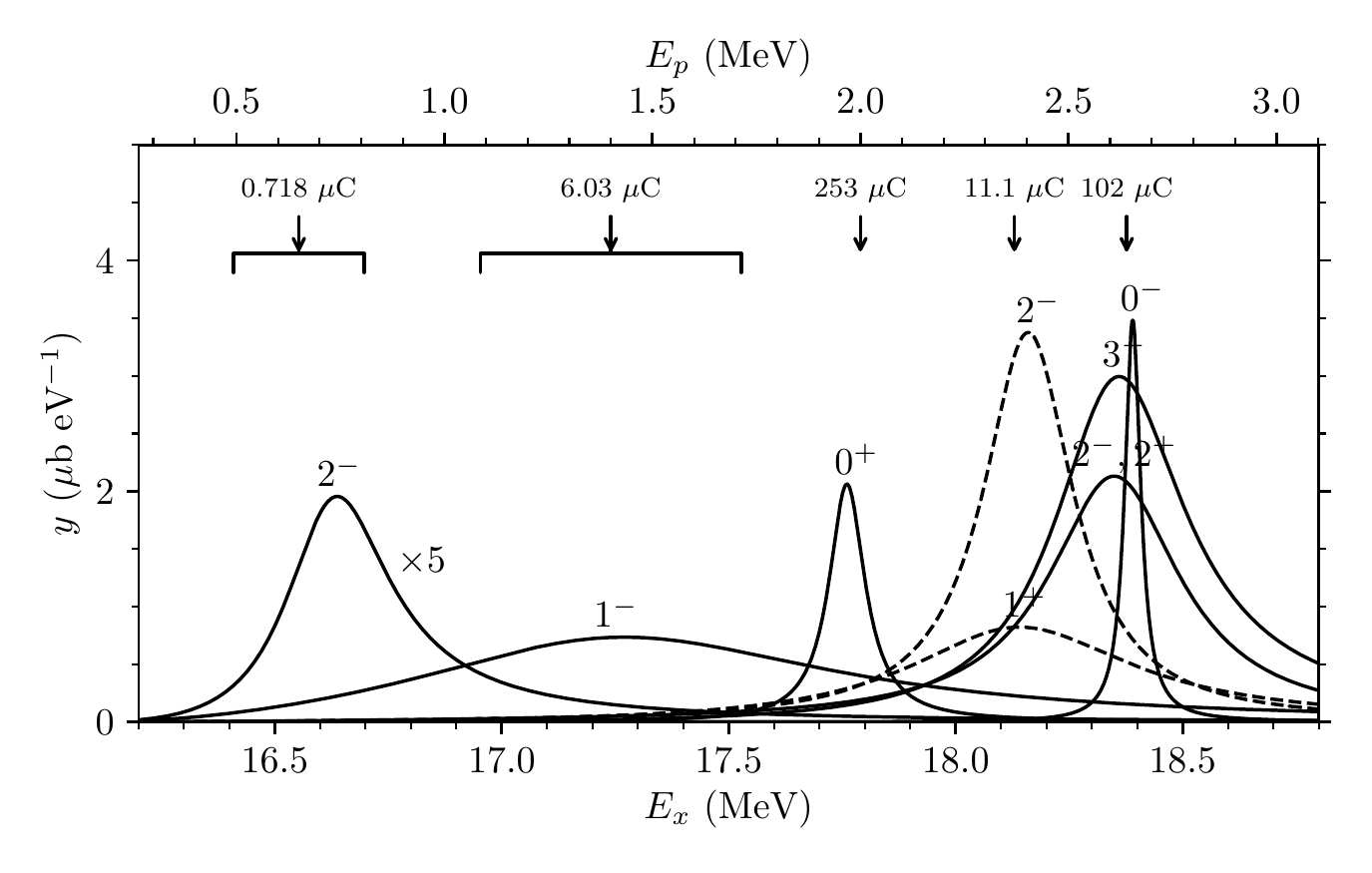}
    \caption{\label{fig:reson-schem}Schematic representation of the known levels in $^{12}$C in the excitation region $E_x = 16.5$--$18.5$~MeV. The quantity $y$, shown on the abscissa, is calculated from Eq.~(\ref{eq:y}). The energies studied in this work are indicated by the arrows. In each case, we give the integrated charge, corrected for the dead time of the data acquisition system. Dashed lines indicate levels that were not found to contribute to the cross sections measured in this work. Note that the 16.62-MeV level has been downscaled by a factor of five for improved display.}
\end{figure}

In Table~\ref{tbl:known-res}, we list known levels in the excitation region $E_x = 16.5$--$18.5$~MeV, which can mediate resonant captures to lower-lying levels at the beam energies investigated in this work. The levels and their properties are obtained from the most recent TUNL compilation~\cite{tunl12} with a few exceptions, as discussed below. %
Fig.~\ref{fig:reson-schem} gives a schematic representation of the levels listed in Table~\ref{tbl:known-res}. The quantity $y$, shown on the abscissa, is calculated from the expression,
\begin{equation}\label{eq:y}
y(E_x) \; = \; 4\pi \lambdabar^2 \omega f(E_x) / \Gamma \; ,
\end{equation}
where the resonance shape is approximated as a Breit-Wigner distribution multiplied by the penetrability for the lowest possible relative orbital angular momentum, 
\begin{equation}\label{eq:bw}
f(E_x) \; = \; \frac{P_{\ell}}{\hat{P}_{\ell}} \times \frac{(\Gamma/2)^2}{(E_x - \hat{E}_x)^2 + (\Gamma/2)^2} \; .
\end{equation}
We note that on resonance, $\sigma_{\gamma,\text{R}} = y \Gamma_{\gamma} \Gamma_p / \Gamma $.

The energies ($\hat{E}_x$) and total widths ($\Gamma$) of the levels listed in Table~\ref{tbl:known-res} are generally well constrained, whereas proton widths ($\Gamma_p$) are either missing or quoted without uncertainties. Proton widths have typically been determined by subtracting the $\alpha$ widths ($\Gamma_{\alpha_0}$, $\Gamma_{\alpha_1}$) from the total width. In particular, $\Gamma_{\alpha_1}$ has been poorly constrained in previous experiments due to the complex $3\alpha$ correlations in this channel~\cite{segel1965}, and therefore the proton widths should be used with some caution. Also, the possibility should not be discounted that the excitation region $16.5$--$18.5$~MeV contains broad $T=0$ levels with large $\alpha$ widths ($\Gamma_{\alpha} > 1$~MeV), which have not been clearly resolved in previous studies.
\begin{table}[b]
{\footnotesize
\begin{minipage}{\linewidth}
\renewcommand{\thefootnote}{\alph{footnote}}
\centering
\caption{\label{tbl:known-res}Known levels in $^{12}$C between 16.5~MeV and 18.5~MeV. Properties obtained from Ref.~\cite{tunl12} with a few exceptions, as discussed in the text. Values in parentheses indicate uncertain assignments.}
\begin{tabular}{ccccc}
\toprule 
$\hat{E}_x$~(MeV) &  $\Gamma$ (keV)  &  $\Gamma_p$ (keV)  &  $J^{\pi}$  &  $T$  \\     
\midrule
16.62(5)  &   280(28)  &   150  &   $2^-$      &  1 \\
17.23     &  1150      &  1000  &   $1^-$      &  1 \\
17.768    &    96(5)   &    76  &   $0^+$      &  1 \\
18.13     &   600(100) &     -  &  ($1^+$)     &  (0) \\
18.16(7)  &   240(50)  &     -  &  ($2^-$)     &  (0) \\
18.35(5)  &   350(50)  &    68  &   $3^-$      &  1 \\
18.35(5)  &   350(50)  &     -  &   $2^-,2^+$  &  $0+1$ \\
(18.39)   &    42      &    33  &   $0^-$      &  (1)\\
\bottomrule   
\end{tabular}
\end{minipage}
}
\end{table}

We proceed by briefly reviewing the available data for each of the levels in Table~\ref{tbl:known-res}. Unless otherwise stated, the data is taken directly from the most recent TUNL compilation~\cite{tunl12}.

\paragraph{16.62, 2$^-$}
The properties of this level are well established, although the precision of $\Gamma_p$ is unclear. The level is clearly observed in $(p,p)$, $(p,\alpha_1)$, and $(p,\gamma_1)$, as established already in the 1950s and 1960s, {\it e.g.}, Refs.~\cite{dearnaley1957, segel1965}. There is also compelling evidence for smaller $\gamma$ branches to the ground state and the \level{12.71}{1}{+} and \level{15.11}{1}{+} levels~\cite{zijderhand1990}, but since the excitation functions were not measured the evidence is not conclusive. 

\paragraph{17.23, 1$^-$}
Owing to its large width, the level is not easily resolved. It is most clearly seen in $(p,\gamma_0)$~\cite{segel1965}, while its precise contribution to $(p,p)$, $(p,\alpha_0)$, and $(p,\alpha_1)$ remains somewhat uncertain. There is compelling evidence for smaller $\gamma$ branches to the \level{4.44}{2}{+}, \level{7.65}{0}{+}, \level{12.71}{1}{+}, and \level{15.11}{1}{+} levels~\cite{zijderhand1990}, but since excitation functions were not measured the evidence is not conclusive. 

\paragraph{17.76, 0$^+$}
The level is seen very clearly in $(p,p)$, $(p,\alpha_0)$, and $(p,\gamma_{12.71})$. The total width has been determined rather accurately by Hanna {\it et al.} and the proton width appears reliable. The level energy of 17.768~MeV was determined from the centroid of the resonance peak in the $(p,\alpha_0)$ spectrum of Ref.~\cite{munch2020}.

\paragraph{18.13, 1$^+$}
Evidence for the existence of this level comes from a single study of $(p,\gamma_{15.11})$~\cite{suffert1972}. There are no constraints on the proton width and the spin-parity and isospin assignments are not conclusive.  
%there is no information about its proton width. However, it is possible to obtain some rough constraints from the lack of clear observation of the state in $(p,p)$ and $(p,\alpha_1)$. Recent $(p,p)$ data (Chiari, Kokkoris) suggests $\sigma_p \lesssim 100$~mb and available $(p,\alpha_1)$ data similarly suggest $\sigma_{\alpha_1} \lesssim 100$~mb, both as conservative upper limits. This leads to the estimate $\Gamma_p / \Gamma \lesssim 0.28$.  

\paragraph{18.16, 2$^-$}
Evidence for the existence of this level also comes from a single study, in this case of $(p,d)$~\cite{lewis1987}. There are no constraints on the proton width and the spin-parity and isospin assignments are not conclusive. It was suggested in Ref.~\cite{lewis1987} that the \level{18.16}{2}{-} and \level{18.13}{1}{+} levels might be one and the same level. Indeed, a spin-parity assignment of $2^-$ appears compatible with the data of Ref.~\cite{suffert1972}. In the TUNL compilation~\cite{tunl12}, the two levels are assumed to be one and the same, but here the 1$^+$ spin-parity assignment of Ref.~\cite{suffert1972} is preferred, while the level energy and width is taken from Ref.~\cite{lewis1987}. However, the very different widths reported in the two studies contradict a single-level interpretation. Therefore, we assume the resonances reported in Refs.~\cite{suffert1972, lewis1987} to correspond to distinct levels. 
%There is no information about its proton width. Using the above arguments, we estimate $\Gamma_p / \Gamma \lesssim 0.14$.

\paragraph{18.35, 3$^-$ \& 2$^-$,2$^+$}
A multidude of experimental probes provide evidence for the existence of at least two, if not three, levels at 18.35~MeV, cf.\ the discussion in Ref.~\cite{Neuschaefer:1983lr}. One of these levels, which is observed both in the spectra of $(p,\alpha_0)$ and $(p,\alpha_1)$ and in the excitation curves of $(p,\gamma_0)$, $(p,\gamma_1)$, and $(p,\gamma_{9.64})$, has been firmly assigned as $3^-$ and isospin $T=1$, with additional evidence to support this assignment coming from $(e,e^{\prime})$ and $^{11}$B$(d,n\alpha_0)$ data~\cite{Neuschaefer:1983lr}. %
On the other hand, $(p,p^{\prime})$ and $(\pi,\pi^{\prime})$ data provide substantial evidence for the presence of an isospin-mixed $2^-$ level at 18.35~MeV with a width similar to that of the $3^-$ level, while $(\alpha,\alpha^{\prime})$ data suggest a $2^+$ level at this energy with isospin $T=0$~\cite{kiss1987}. %
$\gamma$ rays to the \level{12.71}{1}{+} and \level{15.11}{1}{+} levels have also been observed at this energy~\cite{zijderhand1990}, but in the absence of yield-curve measurements they cannot be attributed to the 18.35-MeV level(s) with certainty. %
Given the complicated situation with two or possibly three overlapping levels, the widths quoted in Table~\ref{tbl:known-res} should be used with some caution.
 
\paragraph{18.39, 0$^-$}
The level has only been observed in $(p,p^{\prime})$. Its spin-parity assignment appears firm although it is based solely on cross-section arguments~\cite{segel1965}, while the isospin remains unknown. The total width and proton width both appear reliable.

\subsection{Partial $\gamma$ widths}

In the following, we provide a resonant interpretation of the observed capture cross 
sections that ignores the sub-dominant direct-capture component, {\it i.e.}, $\sigma_{\gamma} \approx \sigma_{\gamma,\textrm{R}}$. With this approximation, partial $\gamma$-decay widths can be deduced directly from Eq.~\ref{eq:resonant}. For those levels where the proton width is unknown, we adopt $\Gamma_p = \Gamma$. This effectively renders the $\gamma$-ray widths deduced for these levels lower limits. For the purpose of estimating off-resonance contributions, we adopt the resonance shapes shown in Fig.~\ref{fig:reson-schem}, taking into account the energy-dependence of the $\gamma$-ray transition rate. We discuss the energy bins I--VII separately, starting with the lowest-energy bin. The deduced $\gamma$-ray widths are summarized in Table~\ref{tbl:gamma-widths}.

\begin{table*}[h]
\begin{minipage}{\linewidth}
{\footnotesize
\renewcommand{\thefootnote}{\alph{footnote}}
\centering
\caption{Transitions observed in the present work. The direct capture component was not considered in the derivation of the partial $\gamma$-ray widths ($\Gamma_{\gamma}$). For those initial levels where the proton width ($\Gamma_p$) is unknown, the derived $\gamma$-ray widths are lower limits. Uncertainties on $\Gamma$ and $\Gamma_p$ have not been taken into account in the estimation of the uncertainty on $\Gamma_{\gamma}$.}
\label{tbl:gamma-widths}
%%\begin{center}%
\begin{tabular*}{\linewidth}{@{\extracolsep{\fill}}cccc *{2}{d{1.6}} }
\toprule 
$E_p$~(MeV)  &  Final level  &  Initial level\footnotemark[1]  &  $ML$  &  \mc{$\Gamma_{\gamma}$~(eV)}  &  \mc{$\Gamma_{\gamma}$~(W.u.)} \\
\midrule
\multirow{2}{*}{2.00, 2.64}  &  \multirow{2}{*}{9.64, $3^-$}   &  18.35, $3^-$  &  $M1$  &  4.7(13)         &  0.34(10)\footnotemark[2]   \\ 
                     &                                &  18.35, $2^-$  &  $M1$  &  1.3(4)          &  0.095(26)\footnotemark[2]  \\
%                     &                                &  18.35, $2^+$  &  $E1$  &  1.3(4)          &  0.0055(15) \\  
\midrule
%0.65                   &  10.84, $1^-$                   &  16.62, $2^-$  &  $M1$  &  \mc{0.28--4.0\footnotemark[3]}  &  \mc{0.070--1.00\footnotemark[3]} \\  
%\midrule
2.00                   &  10.84, $1^-$                   &  17.76, $0^+$  &  $E1$  &  \mc{1.11(21)\footnotemark[3]}  &  \mc{0.0093(18)\footnotemark[3]} \\  
\midrule
\multirow{2}{*}{2.64}  &  \multirow{2}{*}{10.84, $1^-$}  &  18.35, $2^-$  &  $M1$  &  0.75(20)        &  0.086(23)\footnotemark[2]  \\
%                     &                                &  18.35, $2^+$  &  $E1$  &  0.75(20)        &  0.0050(13) \\ 
                     &                                &  18.39, $0^-$  &  $M1$  &  0.80(21)        &  0.090(24)\footnotemark[2]  \\  
\midrule
2.00                   &  11.83, $2^-$                   &  17.23, $1^-$  &  $M1$  &  4.4(14)\footnotemark[4]  &  1.4(4)\footnotemark[4]  \\ 
\midrule
2.00                   &  12.71, $1^+$                   &  17.76, $0^+$  &  $M1$  &  \mc{8.7(19)}  &  \mc{3.2(7)}  \\ 
\midrule
\multirow{2}{*}{2.64}  &  \multirow{2}{*}{12.71, $1^+$}  &  18.35, $2^-$  &  $E1$  &  0.77(28)        &  0.012(4)  \\ 
                     &                                &  18.39, $0^-$  &  $E1$  &  0.81(3)         &  0.013(5)  \\  
\bottomrule   
\end{tabular*}
%%\end{center}
\footnotetext[1]{In those cases, where several initial levels can account for the observed feeding of the final level, all initial levels are given, and the widths are computed assuming that each transition accounts for the full cross section.}
\footnotetext[2]{Some evidence for a smaller contribution from an isoscalar $E1$ transition from the \level{18.35}2{+} level.}
%\footnotetext[3]{Based on the observation of a single event.}
\footnotetext[3]{A substantial contribution from an $M1$ transition from the \level{17.23}{1}{-} level cannot be ruled out.}
\footnotetext[4]{Contributions from transitions from the \level{16.62}{2}{-} and \level{18.13}{1}{+} levels cannot be ruled out.}
}
\end{minipage}
\end{table*}

\paragraph{I)} The yield in the lowest-energy bin is attributed entirely to $p+{}^{10}\text{B}\rightarrow 2\alpha + {}^3\text{He}$, as confirmed by separate measurements performed on an isotope-enriched $^{10}$B target.

\paragraph{II)} At $E_p=2.64$~MeV, the \level{9.64}{3}{-} level is observed very clearly in the $^8$Be$_{\text{gs}}$ channel. The inferred cross section is somewhat smaller than that of Hanna {\it et al.}~\cite{hanna1982}, but consistent within uncertainties. The cross section may be accounted for by isovector $M1$ transitions from the negative-parity levels at 18.35~MeV. An isoscalar $E1$ transition from the $2^+$ level cannot by itself account for the full cross section, as this would require a strength of 0.0055(15)~W.u., exceeding the upper limit of 0.002~W.u.\ recommended for such transitions~\cite{endt1993}. %
However, it was noted by Hanna {\it et al.}\ that the angular distribution of the $\gamma$ ray to the \level{9.64}{3}{-} level is suggestive of mixing between two opposite parity levels, which provides some evidence for a sub-dominant contribution from the $2^+$ level. The two events observed in the $^8$Be$_{\text{exc}}$ channel may be attributed to $\alpha$ decay of the \level{9.64}{3}{-} level via the ghost of the $^8$Be ground state, which has been estimated to account for 2\% of the $\alpha$-decay intensity~\cite{alcorta12}. In Table~\ref{tbl:gamma-widths}, we give the widths required for each of the two candidate transitions to produce the full observed cross section. The width of $4.7(12)$~eV obtained for the \level{18.35}{3}{-}$\rightarrow\;$\level{9.64}{3}{-} transition agrees within uncertainties with the less precise width of $5.7(23)$~eV reported by Hanna {\it et al.}~\cite{hanna1982}. Another estimate of this width can be obtained by combining $\Gamma_{\gamma_1}=3.2(10)$~eV from Ref.~\cite{segel1965}, with the intensity ratio $I_{\gamma_{9.64}} / I_{\gamma_1} = 0.68$ from Ref.~\cite{zijderhand1990} measured at $\theta=55^{\circ}$. This yields  $\sim 2.2$~eV, in reasonable agreement with our value and that of Hanna {\it et al.} Finally, we note that the cross section at $E_p=2.00$~MeV is consistent with feeding of the \level{9.64}{3}{-} level via the low-energy tails of the 18.35-MeV levels. 

\paragraph{III)} Feeding to the \level{10.84}{1}{-} level is observed both at $E_p=2.00$~MeV and 2.64~MeV. At the lower proton energy, where the level is seen very clearly, the cross section is most readily accounted for by an isovector $E1$ transition from the \level{17.78}{0}{+} level with a strength of $0.0128(25)$~W.u., which is typical for such transitions in light nuclei~\cite{endt1993}. An isovector $M1$ transition from the broad \level{17.12}{1}{-} level is also a possibility, although the short measurements performed at $E_p\sim 1.4$~MeV and 2.37~MeV indicate that such a transition could not be the dominant contribution at $E_p=2.00$~MeV. Assuming this were the case, we would expect to observe 3.0--3.5 events at $E_p\sim 1.4$~MeV whereas only one event was observed ($1.8\sigma$ discrepancy~\cite{rolke2005}) and 2.5--3.0 events at $E_p=2.37$~MeV whereas only one event was observed ($1.3\sigma$ discrepancy). %
(We note that there is a slight mismatch between the energies of the two observed events, $E_x = 11.15$~MeV and 11.25~MeV, respectively, and the energy of the \level{10.84}{1}{-} level, leading to some uncertainty in their interpretation.) %
The feeding observed in the $^8$Be$_{\text{exc}}$ channel may be attributed to $\alpha$ decay of the \level{10.84}{1}{-} level via the ghost of the $^8$Be ground state, which has been esimated to account for 8\% of the $\alpha$-decay intensity~\cite{alcorta12}.

At $E_p=2.64$~MeV, where the feeding of the \level{10.84}{1}{-} level is less pronounced, the cross section is consistent with isovector $M1$ transitions from the \level{18.35}{2}{-} level or the \level{18.38}{0}{-} level, while an isoscalar $E1$ transition from the \level{18.35}{2}{+} level is ruled out because the required strength exceeds the recommended upper limit for such transitions~\cite{endt1993}. %

Finally, we note that in the short measurement performed at $E_p\sim 0.65$~MeV, a single event was detected in the $^8$Be$_{\text{gs}}$ channel. This event had an energy consistent with that of the \level{10.84}{1}{-} level and could be accounted for by an isovector $M1$ transition from the \level{16.62}{2}{-} level with a strength of 0.070--1.00~W.u., which is typical for transitions of this kind in light nuclei~\cite{endt1993}.
%The width of 0.54 eV for the 16.62->12.71 transition (deduced from \cite{zijderhand1990}) is consistent with its non-observation in our experiment. Conversely, the transition to the 10.84 MeV state seen here (with 1 count) with a width of 0.2--3 eV, would have been missed by \cite{zijderhand1990} because the 10.84 MeV state is wide.

\paragraph{IV)} At $E_p=2.00$~MeV, a peak occurs in the cross section at $E_x\sim 11.8$~MeV in both the $^8$Be$_{\text{gs}}$ and $^8$Be$_{\text{exc}}$ channel. While the \level{11.83}{2}{-} level provides a natural explanation for the peak in the $^8$Be$_{\text{exc}}$ channel, this level cannot account for the peak in the $^8$Be$_{\text{gs}}$ channel, which requires a level of natural parity. At $E_p=2.64$~MeV, strength is observed in both channels, but there is no clear indication of a peak at 11.8~MeV, suggesting that the \level{11.83}{2}{-} level only makes a minor contribution to the cross section at this proton energy. %
The feeding to the \level{11.83}{2}{-} level at $E_p=2.00$~MeV is most naturally accounted for by an isovector $M1$ transition from the \level{17.23}{1}{-} level with a strength of 1.4(4)~W.u. An $E1$ transition from the \level{18.13}{1}{+} level could also be contributing, but cannot account for the entire feeding. If it did, we would expect to observe 43(7) events at $E_p=2.64$~MeV whereas only 19 events were observed in the $^8$Be$_{\text{exc}}$ channel (3.0$\sigma$ discrepancy). An isovector $M1$ transition from the \level{16.62}{2}{-} level provides yet another potential feeding mechanism, inconsistent only at the level of 2.2$\sigma$ with the low-statistics data collected at $E_p=0.65$~MeV, but requires a rather large strength of 5.4~W.u.\ to account for the entire cross section. %

We now turn to the observation of a peak-like structure at $E_x\sim 11.8$~MeV in the $^8$Be$_{\text{gs}}$ channel at $E_p=2.00$~MeV, which is intriguing since no narrow levels with natural-parity are known to exist at this energy in $^{12}$C. Unfortunately, the data provide few constraints on the quantum numbers of the level, only ruling out spins $J \geq 4$: Feeding of a $0^+$ level can be accounted for by an $M1$ transition from the \level{17.23}{1}{-} level; feeding of a $1^-$ level by $M1$ transitions from the \level{16.62}{2}{-} and \level{17.23}{1}{-} levels or an $E1$ transition from the \level{17.76}{0}{+} level; feeding of a $2^+$ level by $E1$ transitions from the \level{16.62}{2}{-} and \level{17.23}{1}{-} levels; and feeding of a $3^-$ level by an $M1$ transition from the \level{16.62}{2}{-} level. In all cases, the required strengths are within expectations for light nuclei~\cite{endt1993} and consistent with the cross sections measured at the other beam energies. %

The cross section measurement at $E_p=2.64$~MeV also provides limited insight into the properties of the final level: Any of the spin-parities $1^-$, $2^+$, $3^-$, and $4^+$ can be accounted for by more than one transition. Only a $0^+$ assignment seems improbable as it requires an isoscalar $E2$ transition from the \level{18.35}{2}{+} level with a rather large strength of 18(6)~W.u.

\paragraph{V)} Feeding to the \level{12.71}{1}{+} level is observed very clearly at $E_p=2.00$~MeV and also at $E_p=2.64$~MeV albeit less clearly. Some cross section is also observed in the $^8$Be$_{\text{gs}}$ channel which cannot be accounted for by the \level{12.71}{1}{+} level. %
The cross section obtained at the lower proton energy is about two times larger than that of Hanna {\it et al.} Even considering the substantial uncertainty on the value of Hanna {\it et al.}, the discrepancy is significant. However, we note that Hanna {\it et al.} relied on the $\gamma_1$ yield reported by Segel {\it et al.}~\cite{segel1965} for normalizing their data, and this yield disagrees with other measurements by up to 50\% as discussed in Ref.~\cite{segel1965}. Also, the $(p,\alpha_0)$ cross section reported by Segel {\it et al.}\ has recently been found to be underestimated by a factor of $1.50^{+0.15}_{-0.11}$~\cite{munch2020}. Taken together, these observations cast doubt on the accuracy of the normalization of the measurements of Hanna {\it et al.} indicating a potential $\sim 50$\% underestimation.

As already noted by Hanna {\it et al.}, the feeding to the \level{12.71}{1}{+} level at $E_p=2.00$~MeV can be accounted for by a rather strong isovector $M1$ transition from the \level{17.76}{0}{+} level. Indeed, the feeding cannot be accounted for in any other way. Adopting our larger cross section, the required strength is 4.4(9)$^{+2.2}_{-1.1}$~W.u., making the transition one of the strongest of its kind~\cite{endt1993}. %
The feeding observed at $E_p=2.64$~MeV cannot be accounted for by the high-energy tail of the \level{17.76}{0}{+} level, but requires an isovector $E1$ transition from either the \level{18.35}{2}{-} or the \level{18.39}{0}{-} level. %

While the cross section observed in the $^8$Be$_{\text{gs}}$ channel is relatively small, it is of substantial interest since no natural-parity levels are known at $E_x\sim 12.7$~MeV. There is, however, evidence for a broad ($\Gamma=1.7$~MeV) level at $E_x=13.3$~MeV with spin-parity $4^+$, the low-energy tail of which could potentially account for the observed cross section. This possibility will be explored further below. %

\paragraph{VI)} The excitation region $E_x = 13$--$14.5$~MeV is known to contain a $4^-$ level at 13.32~MeV, which decays entirely via the $^8$Be$_{\text{exc}}$ channel, and a $4^+$ level at 14.08~MeV, which decays predominantly via the $^8$Be$_{\text{exc}}$ channel (78\%). Recently, evidence has been found for a very broad ($\Gamma = 1.7$~MeV) $4^+$ level at 13.3~MeV. We observe relatively little feeding into this region, consistent with the expected inhibition of $\gamma$ transitions that require large changes in spin. We note that the factor of $\sim 4$ enhancement of the cross section in the $^8$Be$_{\text{exc}}$ channel compared to the $^8$Be$_{\text{gs}}$ channel appears consistent with the known decay properties of the known levels, especially if the broad 13.3-MeV level is assumed to have a substantial decay component to the $^8$Be ground state. %

Only isovector $E1$/$M1$ transitions from the \level{18.35}{3}{-} level can account for the feeding to the $4^{\pm}$ levels. However, this mechanism should produce a factor of $\sim 15$ enhancement of the cross section at $E_p=2.64$~MeV relative to 2.00~MeV, which is not observed. The discrepancy could potentially be reduced somewhat if the asymmetric shape of the 18.35-MeV level were taken into account, but it seems unlikely that this can fully explain the discrepancy. This suggests two possibilities: some of the cross section observed at the lower proton energy is to be attributed to {\it (i)} feeding {\it to} an unknown natural-parity level with $E_x \sim 13$--14~MeV and $J \leq 2$, or {\it (ii)} feeding {\it from} an unknown level with $E_x \sim 17$--18~MeV and $J \geq 2$.

\paragraph{VII)} At both $E_p=2.00$~MeV and 2.64~MeV, we observe substantial feeding to the excitation region above 14.5~MeV, especially in the $^8$Be$_{\text{exc}}$ channel. It seems natural to ascribe the majority of this cross section to the broad level at 15.44~MeV, tentatively assigned as $2^+$ although $0^+$ has also been proposed, but the feeding to this level is problematic:

Adopting the $2^+$ assignment, the cross section observed at $E_p=2.00$~MeV can only be accounted for by an isovector $E1$ transition from the \level{17.23}{1}{-} level, but we dismiss this possibility because the required strength of 2.5(8)~W.u.\ exceeds the recommended upper limit of 0.5~W.u.~\cite{endt1993} by a factor of five. (Transitions from the levels above 18~MeV can be dismissed because they overpredict the cross section at $E_p=2.64$~MeV.) %
Adopting instead the $0^+$ assignment, the conclusion is the same: no transition from any of the known levels can account for the observed feeding while conforming to the recommended upper limits of Ref.~\cite{endt1993}.

At $E_p=2.64$~MeV, the feeding can be accounted for by a rather strong $M1$ transition from the \level{18.35}{2}{-} level with a strength of (at least) 0.29(9)~W.u., but only if the $2^+$ assignment is adopted for the 15.44-MeV level.

\section{Summary and conclusions}
We summarize our findings as follows: The $^{11}\text{B}(p,3\alpha)\gamma$ cross sections measured at $E_p=2.00$~MeV and 2.64~MeV give clear evidence of feeding to the four known levels \level{9.64}{3}{-}, \level{10.84}{1}{-}, \level{11.83}{2}{-}, and \level{12.71}{1}{+}, but by themselves these levels cannot fully account for the observed cross sections. In particular, we find evidence for feeding to a natural-parity level near $E_x \sim 11.8$~MeV. Evidence for natural-parity strength in this region was also found in a previous study of the $\gamma$ de-excitations of the \level{16.11}{2}{+} level \cite{laursen2016_2} and in studies of the $\beta$ decays of $^{12}$B and $^{12}$N \cite{hyldegaard2010}.

The feeding to the \level{9.64}{3}{-}, \level{10.84}{1}{-}, \level{11.83}{2}{-}, and \level{12.71}{1}{+} levels can be explained in terms of isovector $M1$ and $E1$ transitions from the known levels above the $p+{}^{11}$B threshold. The transitions proposed to account for the feeding to the \level{10.84}{1}{-}  and \level{12.71}{1}{+} levels at $E_p=2.64$~MeV are of some interest, as they provide evidence for significant $T=1$ admixture in the \level{18.35}{2}{-} level and/or the \level{18.39}{0}{-} level. It is also worth noting that the larger and more precise width obtained for the \level{17.76}{0}{+}$\rightarrow\;$\level{12.71}{1}{+} transition makes this one of the strongest $M1$ transition in any nucleus~\cite{endt1993}. 

Higher-statistics measurements at $E_p = 0.65$~MeV and 1.4~MeV would be highly desirable to confirm the tentative observation of $M1$ transitions from the \level{16.62}{2}{-} and \level{17.23}{1}{-} levels, both feeding into the \level{10.84}{1}{-} level. Such measurements would also yield improved constraints on the spin-parity of the natural-parity level observed at $E_x \sim 11.8$~MeV. For these studies, it could prove advantageous to adopt a detector geometry similar to that of Ref.~\cite{laursen2016_2}, which allows significantly larger beam currents at the cost of a substantial reduction of the detection efficiency in the $^8$Be$_{\text{exc}}$ channel.

The interpretation of the feeding observed into the excitation region above 13~MeV remains unclear, especially at $E_p=2.00$~MeV where the measured cross section could not be explained in terms of transitions between known levels. Here, too, additional measurements would be desirable.

An analysis of new complete-kinematics data on the $^{11}$B$(p,3\alpha)$ reaction, currently in progress, will provide an improved understanding of the $\alpha_1$ channel. This, together with a multi-channel $R$-matrix analysis that includes recent data on $(p,p)$ and $(p,\alpha_0)$ as well as existing data on other channels, should lead to an improved understanding of the excitation region $E_x\sim \;$16--18 MeV, which may require revision of some of the conclusions drawn from the present study.

While theoretical estimates suggest the resonant capture component to be dominant, the direct 
capture component is not negligible and could in some instances make a substantial contribution 
to the observed cross section. Such direct contributions were not considered in the derivation of the partial $\gamma$-ray widths given in Table~\ref{tbl:gamma-widths}. Improved theoretical calculations of the direct component would be of significant interest. Theoretical calculations of the radiative widths deduced in this work would also be of interest. 

Finally, we remark that the $2^+\rightarrow 0^+$ and $4^+\rightarrow 2^+$ transitions in $^8$Be contribute only at the sub-nb level to the cross sections measured in this work, and hence can be safely ignored~\cite{datar2005}.

\section*{Acknowledgements}

We would like to thank Folmer Lyckegaard for manufacturing the target. %
This work has been supported by the European Research Council under ERC 
starting grant LOBENA, No. 307447. %
OSK acknowledges support from the Villum Foundation through 
Project No.\ 10117.

%%%\bibliography{refs}

\end{document}